\documentclass[twoside,12pt]{article}
\usepackage{epsfig}
\usepackage{graphicx}
\def\Journal#1#2#3#4{{#1} {#2} (#4) #3 }
\def\AOP{{\em Ann.\ Phys.}}
\def\EPL{{\em Europhys.\ Lett.}}
\def\EPJC{{\em Eur.\ Phys.\ J.} C.}

\def\JPG{{\em J.\ Phys.} G}

\def\PLB{{\em Phys.\ Lett.} B}

\def\PRL{{\em Phys.\ Rev.\ Lett.}}

\def\PRD{{\em Phys.\ Rev.} D}

\def\ZPC{{\em Z.\ Phys.} C}

\newcommand{\be}{\begin{equation}}
\newcommand{\ee}{\end{equation}}
\newcommand{\bea}{\begin{eqnarray}}
\newcommand{\eea}{\end{eqnarray}}

\newcommand{\rse}{\mathcal{R}}
\newcommand{\bes}[1]{j^{#1}_{L_{#1}}}
\newcommand{\han}[1]{h^{(1)#1}_{L_{#1}}}
\newcommand{\pot}[2]{V_{#1#2}^{(L_{#1},L_{#2})}}
\newcommand{\tmat}[2]{T_{#1#2}^{(L_{#1},L_{#2})}}
\newcommand{\One}{1\!\!1}

\newcommand{\dsc}{$D_{s1}(2536)$}
\newcommand{\dsd}{$D_{s1}(2460)$}
\newcommand{\dc}{$D_{1}(2420)$}
\newcommand{\dd}{$D_{1}(2430)$}
\newcommand{\tpz}{${}^{3\!}P_0$}
\newcommand{\tpo}{${}^{3\!}P_1$}
\newcommand{\otpo}{$1\,{}^{3\!}P_1$}
\newcommand{\ttpo}{$2\,{}^{3\!}P_1$}
\newcommand{\osdt}{$1\,{}^{1\!}D_2$}
\newcommand{\otdo}{$1\,{}^{3\!}D_1$}
\newcommand{\spo}{${}^{1\!}P_1$}

\newcommand{\ttdo}{$2\,{}^{3\!}D_1$}
\newcommand{\jpsi}{J\!/\!\psi}

\begin{document}

\title{ \vspace{1cm} Towards Meson Spectroscopy Instead of Bump Hunting}
\author{George Rupp,$^1$ Susana Coito,$^1$ Eef van Beveren$^2$ \\ \\
$^1$Centro de F\'{\i}sica das Interac\c{c}\~{o}es Fundamentais,
Instituto Superior T\'{e}cnico, \\ Technical University of Lisbon,
P-1049-001 Lisboa, Portugal \\[1mm]
$^2$Centro de F\'{\i}sica Computacional,
Departamento de F\'{\i}sica, \\ Universidade de Coimbra,
P-3004-516 Coimbra, Portugal}

\maketitle
\begin{abstract} 
Mesonic resonances are generally observed in data as narrow, moderately
broad, or wide peaks in scattering or production processes. In the eyes of
nearly all experimentalists, any suchlike bump is a true resonance
as soon as its statistical significance exceeds certain minimal
values. However, this simple point of view ignores
possible effects from competing hadronic channels and the opening
of the corresponding thresholds. On the other hand, most theoretical
hadron-model builders consider mesons merely bound states of a quark
and an antiquark, or of more exotic combinations sometimes involving
valence gluons as well. Also the latter description is much too
naive, since considerable mass shifts or even the dynamical
generation of extra states due to unquenching are equally ignored.

In the present paper, a largely empirical yet very successful
approach to meson spectroscopy is revisited, in which all the above
phenomena can be accounted for non-perturbatively, with concrete examples
of some enigmatic mesonic states described in detail. First, the $X(4260)$
charmonium enhancement is argued to be a non-resonant structure
resulting from depletion effects due to competing channels and
resonances. Then, the $X(3872)$ charmonium-like meson  is described as a
unitarised $J^{PC}=1^{++}$ $c\bar{c}$ state. Also, the unusual pattern
of masses and widths of the open-charm axial-vector mesons \dc, \dd, \dsc,
and \dsd\ is shown to follow from highly non-perturbative coupled-channel
and mixing effects. Finally, first indications of a very light scalar boson
are presented, on the basis of published BABAR data.
\end{abstract}

\section{Introduction}
Meson spectroscopy is an essential tool in trying to understand both the
confinement and the decay mechanism in QCD. The large volume of experimental
data \cite{PDG2010} on mesonic bound states and resonances, from the 
light-quark sector all the way up to bottomonium, should allow to draw
far-reaching and detailed conclusions on the nature of the confining potential
and how it is affected by the possibility of strong decay, via $q\bar{q}$
pair creation. However, a necessary condition for a systematic analysis of
mesonic data for spectroscopy purposes is the ability to distinguish between
true resonances and enhancements that originate in threshold effects, as well
as inelasticities due to competing channels and true resonances. Especially
in charmonium, several intriguing structures, a couple of which even charged,
have been observed \cite{PDG2010} above the lowest open-charm threshold that
may not be resonances at all \cite{PoSHQL2010p003}. Very recently, also in
bottomonium two charged enhancements were reported \cite{HEPEX11054583},
which happen to lie just a few MeV above the $BB^\ast$ and
$B^\ast B^\ast$ thresholds, respectively. These may be other examples of
non-resonant structures that are just threshold cusps \cite{EPL96p11002}. 
However, see Ref.~\cite{HEPPH11061552} for a different point of view.

On the other hand, there are true resonances, too, that do not show up as clear
bumps in the data, or are not accompanied by a scattering phase shift rising
past $90^\circ$. Examples of the latter case are the light scalar mesons
\cite{ZPC30p615,PLB641p265} $f_0(600)$ \cite{PDG2010} (alias $\sigma$) and 
$K_0^\ast(800)$ (alias $\kappa$) \cite{PDG2010}. What makes their experimental
observation so difficult is a combination of factors, namely their very large
widths, slowly rising phase shifts due do an Adler zero slightly below the
lowest threshold
(at $\approx\!m_\pi^2/2$ for $\sigma$, $\approx\!m_K^2-m_\pi^2/2$ for
$\kappa$), and partly overlapping resonances with the same quantum numbers
($f_0(980)$ for $\sigma$, $K_0^\ast(1430)$ for $\kappa$).

Thus, non-resonant enhancements and hard to observe resonances complicate
the task of meson spectroscopists in trying to systematically describe the
available data. However, an additional and often overlooked difficulty is
the inevitable mass shift of $q\bar{q}$ states, bound by some confining
potential, due to the possibility of quark-pair creation and annihilation,
which gives rise to meson-loop contributions, even for states below all
strong decay threshold. These shifts can, in principle, be as large as the
hadronic decay widths of many mesons, i.e., of the order of hundreds of MeV.
Moreover, in general one is dealing with complex and non-linear energy shifts,
governed by analytic $S$-matrix properties, the widths being generated by the
very same mechanism. In view of such effects, it is hard to understand why so
many model builders immediately resort to exotic configurations, such as
tetraquarks and hybrids, whenever a newly observed mesonic resonance does not
seem to fit in the usual quark potential models. Rather, one should try to
``unquench'' these models, in the same spirit as in modern lattice-QCD
calculations.

In the present short review, we first show why we believe the $X(4260)$ 
enhancement, discovered by the BABAR collaboration, not to be a
resonance, on the basis of
the BABAR data themselves. Next, the Resonance-Spectrum Expansion (RSE) is
briefly revisited, which is the non-perturbative and exactly solvable 
unquenched formalism we employ to describe non-exotic mesonic resonances. The
RSE method is then applied to the $X(3872)$ charmonium state and the
axial-vector charmed mesons \dc, \dd, \dsc, and \dsd. Finally, as a surprising
additional result, we present indications of a so far unobserved very light
scalar boson, from analysing published BABAR data.

\section{Non-resonant charmonium enhancement $X(4260)$}
In 2005, the BABAR Collaboration observed \cite{PRL95p142001} the vector
charmonium enhancement $X(4260)$ in $\pi^+\pi^-\jpsi$ data, which was later
confirmed by other collaborations and also in different production processes,
though not in open-charm channels \cite{PDG2010}. The non-observation of 
OZI-allowed  decays of the $X(4260)$, as well as its awkward mass for most 
potential models, led to various exotic or molecular model explanations (see
Ref.~\cite{PRL105p102001} for some references). However, another very striking
feature of the $X(4260)$ data, viz.\ a conspicuous dip precisely at the mass of
the well-established $\psi(4415)$ \cite{PDG2010} resonance, is usually ignored.
Moreover, none of the other known $c\bar{c}$ resonances show up in the BABAR
data \cite{PRL95p142001}. Nevertheless, this seemingly inexplicable behaviour
of the $X(4260)\to\pi^+\pi^-\jpsi$ data can be understood if one assumes that
they are depleted by competing yet not searched for open-charm decays, which
effect should be most pronounced at the positions of true resonances. Such a
depletion makes sense, as decays into open-charm mesons are OZI-allowed and
therefore should be strongly dominant.

\begin{figure}[htb]
\begin{tabular}{c}
\hspace*{2.3cm}
\includegraphics[scale=0.70,trim=0cm 0.4cm 0cm 0.4cm, clip=true]
{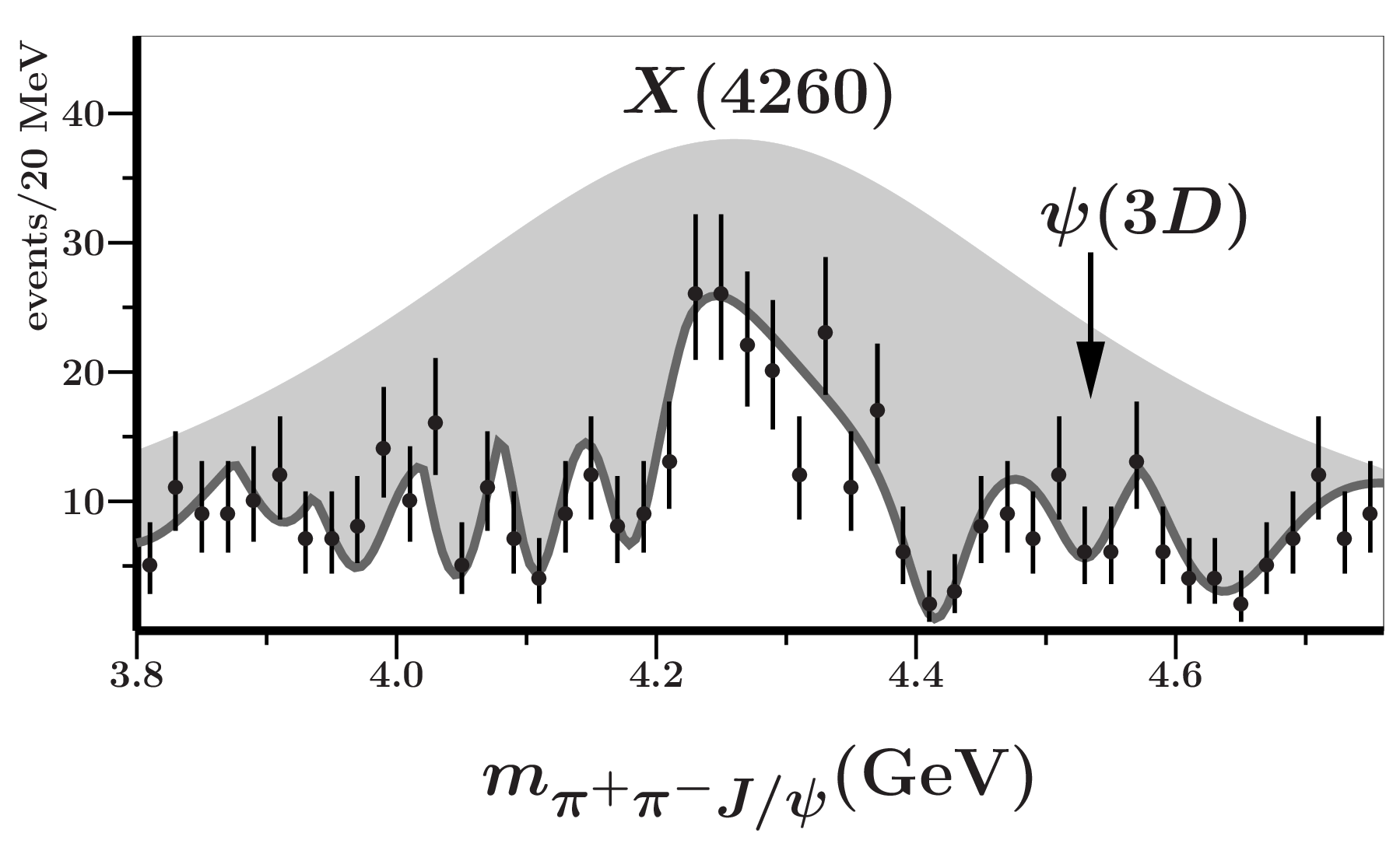}
\end{tabular}
\caption{Depletion of very broad structure (shaded area) by $c\bar{c}$ resonances
and open-charm thresholds, resulting in solid curve. Data are from 
Ref.~\cite{PRL95p142001}; see Ref.~\cite{PRL105p102001} for further details.}
\label{octopsi}
\end{figure}

Reference~\cite{PRL105p102001} elaborated on this idea, by assuming a very
broad, non-resonant structure centred on the $X(4260)$, resulting from an
$f_0(600)$ (and possibly $f_0(980)$) peripheral emission in an OZI-forbidden
production process like $e^+e^-\to c\bar{c}\to (f_0(600)\to\pi^+\pi^-)\jpsi$.
The competing open-charm thresholds $D\bar{D}$, $D\bar{D}^{\ast}$,
$D^{\ast}\bar{D}^{\ast}$, $D_{s}\bar{D}_{s}$, $D_{s}\bar{D}_{s}^{\ast}$,
$D_{s}^{\ast}\bar{D}_{s}^{\ast}$, and $\Lambda_{c}\Lambda_{c}$, as
well as the $c\bar{c}$ resonances $\psi (4040)$, $\psi (4160)$, $\psi (4415)$,
$\psi (3D)$, then deplete this structure, resulting in the data pattern
observed by BABAR, including the $X(4260)$ enhancement, as depicted in
Fig.~\ref{octopsi}.
Note that the $\psi(3D)$ state, at about 4.53 GeV, is not yet listed in the
PDG tables \cite{PDG2010}, but its existence is supported by other data as
well (see Ref.~\cite{PRL105p102001} and references therein, also for more details
on the data parametrisation employed in Fig.~\ref{octopsi}).

\section{Resonance-Spectrum Expansion}
\begin{figure}[htb]
\begin{tabular}{cc}
\resizebox{!}{65pt}{\includegraphics{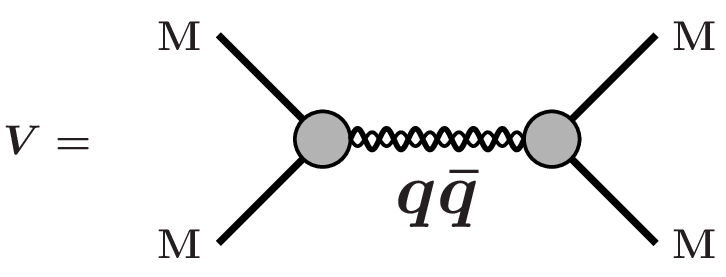}} &
\hspace*{20pt}\resizebox{!}{65pt}{\includegraphics{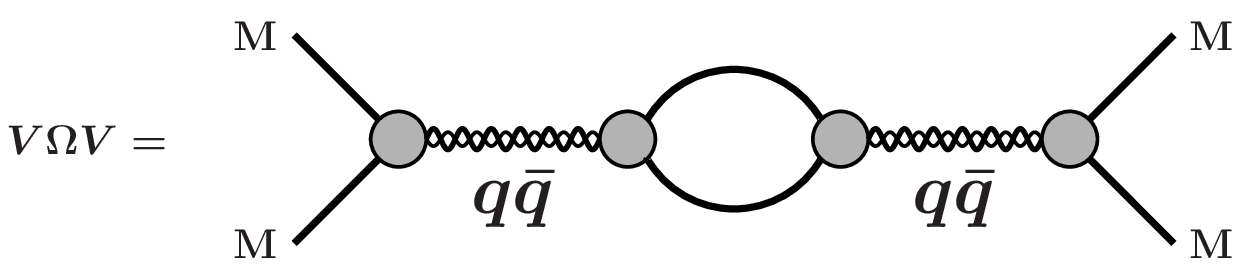}}
\end{tabular}
\caption{Left: Effective meson-meson Born diagram.
Right: corresponding one-loop diagram.}
\label{effective}
\end{figure}

For the description of meson-meson scattering in non-exotic channels and
resonances therein, the Resonance-Spectrum Expansion
\cite{AOP324p1620,AOP323p1215} is a
simple though powerful and only mildly model-dependent formalism. It is based
on the idea that $s$-channel exchanges strongly dominate such scattering
processes, giving rise to an effective meson-meson interaction that is
separable in the relative momenta, as can be seen from Fig.~\ref{effective}.
However, contrary to other effective approaches in which sometimes $s$-channel
seeds are introduced, our intermediate states are described microscopically by a
complete spectrum of bare quark-antiquark systems, thus resembling Regge
propagators \cite{AOP324p1620} (see the wiggly lines in
Fig.~\ref{effective}). The transitions between the $q\bar{q}$ and
meson-meson systems are supposed to take place via \tpz\ pair creation or
annihilation, modelled via a spherical delta function in coordinate space,
which mimics string breaking at a well-defined separation. The resulting
vertex functions, represented by the blobs in the Born and one-loop 
diagrams in Fig.~\ref{effective}, are spherical Bessel functions in momentum
space. Owing to the separability of the effective interaction, the
corresponding $T$-matrix can be solved in closed form. In the general case of
$N_{q\bar{q}}$ $q\bar{q}$ channels coupled to $N_{MM}$ meson-meson channels,
the interaction reads
\begin{equation}
\pot{i}{j}(p_i,p'_j;E) \; = \; \displaystyle\lambda^2r_0\,\bes{i}(p_ir_0)\,
\bes{j}(p'_jr_0)\,\sum_{\alpha=1}^{N_{q\bar{q}}}\sum_{n=0}^{\infty}
\displaystyle\frac{g_i^{(\alpha)}(n) g_j^{(\alpha)}(n)}{E-E_n^{(\alpha)}}
\; \equiv \; \mathcal{R}_{ij}(E)\,\bes{i}(p_ir_0)\,\bes{j}(p'_jr_0)\;,
\label{veff}
\end{equation}
leading to the fully off-energy-shell $T$-matrix 
\begin{equation}
\tmat{i}{j}(p_i,p'_j;E) \; = \; 
-2\lambda^2r_0\,\sqrt{\mu_ip_i\mu'_jp'_j}\:\bes{i}(p_ir_0)\sum_{m=1}^{N}
\rse_{im}(E)\,
\left\{[\One-\Omega\,\mathcal{R}]^{-1}\right\}_{\!mj}\,\bes{j}(p'_jr_0)\;,
\label{tfinal}
\end{equation}
with the loop function
\begin{equation}
\Omega \; = \; -2i\lambda^2r_0\:\mbox{diag}\left(\bes{n}(k_nr_0)
\han{n}(k_nr_0)\right) \; . 
\label{loop}
\end{equation}
Here, $\lambda$ is an overall coupling,
$r_0$ is the average distance for decay via \tpz\ quark-pair creation,
$E_n^{(\alpha)}$ is the discrete energy of the $n$-th recurrence in $q\bar{q}$
channel $\alpha$, $g_i^{(\alpha)}$ is the corresponding coupling to the
$i$th meson-meson channel, $\mu_i$ the reduced mass for this channel, $p_i$ the
off-shell relative momentum, $L_i$ the orbital angular momentum, and
$\bes{i}$ and $\han{j}(k_jr_0)$ the spherical Bessel function and Hankel
function of the first kind, respectively. Note that $\mu_i$, $p_i$, and the
on-energy-shell relative momentum $k_i$ are defined relativistically.
The $S$-matrix is then given by
$S^{(L_i,L_j)}_{ij}(E)=1+2i\tmat{i}{j}(k_i,k_j;E)$. 

\section{$X(3872)$ as a unitarised $1^{++}$ $c\bar{c}$ state}
The $X(3872)$ charmonium-like state was discovered in 2003 by the Belle
Collaboration \cite{PRL91p262001}, as a $\pi^+\pi^-\jpsi$ enhancement in the
decay $B^\pm\to K^\pm\pi^+\pi^-\jpsi$. In subsequent years, it was confirmed
in other experiments, with the additional hadronic decay modes $\rho^0\jpsi$,
$\omega\jpsi$, $D^0\bar{D}^0\pi^0$, and $D^0\bar{D}^{\ast0}$ being seen as
well \cite{PDG2010}. Although the $X(3872)$ decay modes are perfectly
compatible with a \ttpo\ or \osdt\ assignment \cite{PDG2010}, the former
$c\bar{c}$ state is usually thought to be roughly 100 MeV heavier, whereas the
latter is mostly predicted to be 30--80~MeV lighter. Also because of the 
$X(3872)$'s remarkable closeness to the $D^0D^{\ast0}$ threshold, many authors
have proposed exotic or molecular configurations (see Ref.~\cite{EPJC71p1762}
for some references).

\begin{figure}[!hb]
\begin{tabular}{cc}
\hspace*{1cm}
\includegraphics[scale=0.5,trim=0cm -1.6cm 0cm 0cm,clip=true]
{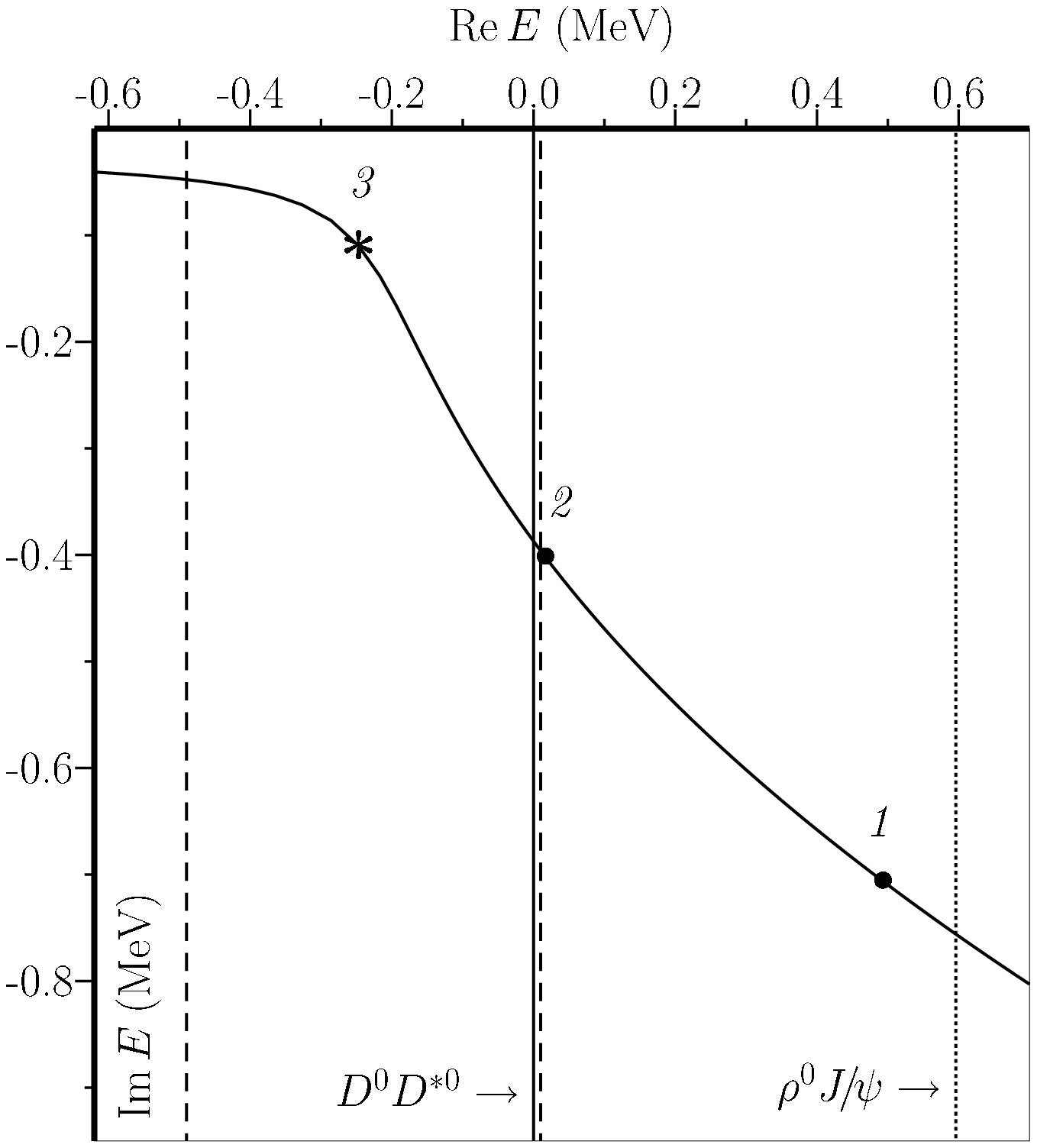} & \hspace*{1cm}
\includegraphics[scale=0.5]{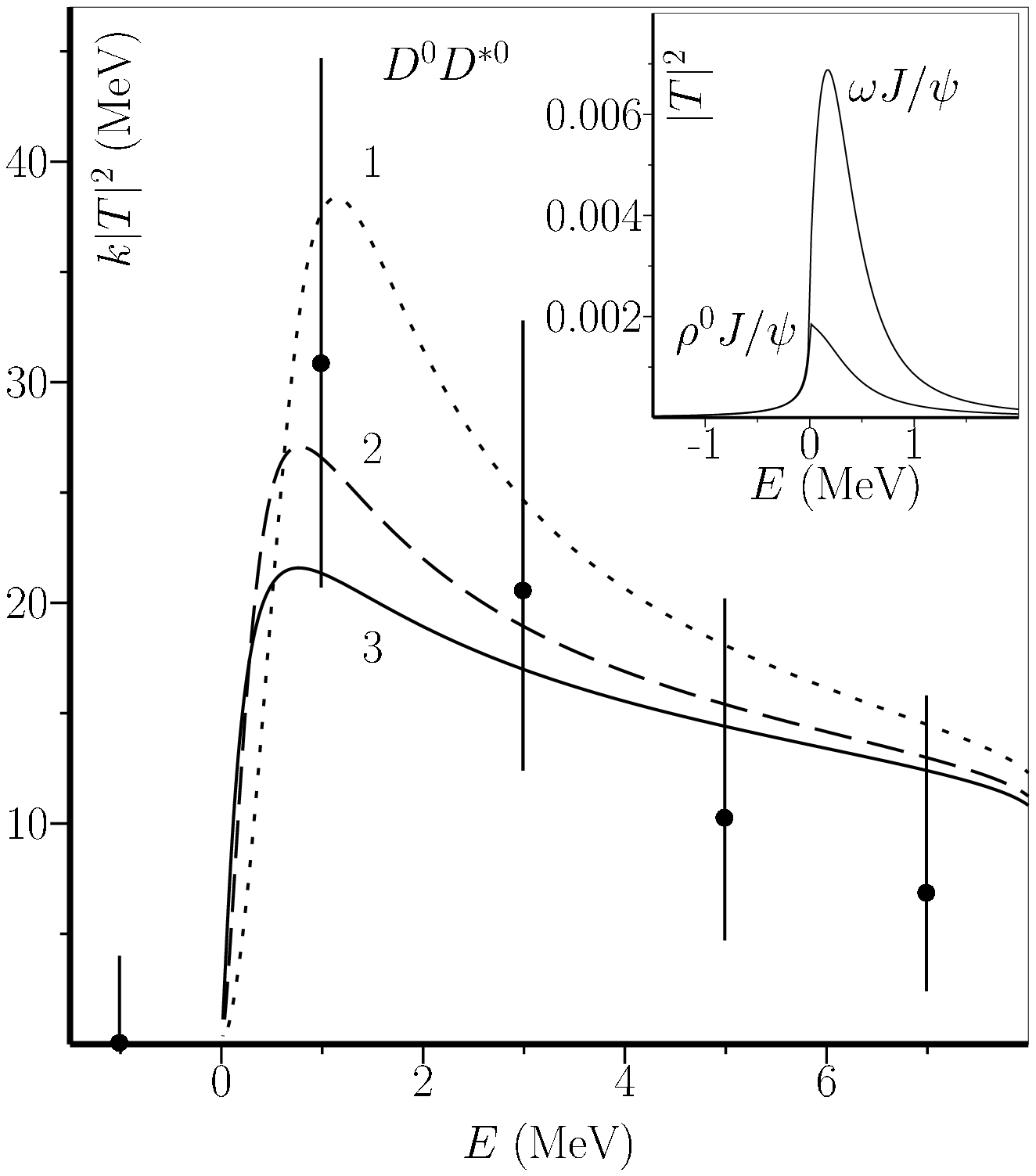}
\end{tabular}
\caption{Left: $X(3872)$ pole trajectory as a function of $\lambda$; dashed
lines delimit allowed \cite{PDG2010} pole range. Right:
$D^0D^{\ast0}$ amplitude with data \cite{PRL91p262001}; inset: relative
comparison of $\rho^0\jpsi$ and $\omega\jpsi$ amplitudes. \vspace*{-3mm}}
\label{x3872}
\end{figure}

However, in Ref.~\cite{EPJC71p1762} we showed that the $X(3872)$ can be very
well described as the first radial excitation of the \otpo\ $\chi_{c1}(3511)$,
i.e., with quantum numbers $J^{PC}=1^{++}$. By coupling the relevant
open and closed open-charm channels to a bare \ttpo\ $c\bar{c}$ state, its
mass gets shifted downwards by roughly 100 MeV, while the corresponding pole
settles almost on top of or slightly below the $D^0D^{\ast0}$ threshold.
In Fig.~\ref{x3872}, 
left-hand plot, we display the $X(3872)$ pole trajectory
as a function of the overall coupling $\lambda$ in the vicinity of the
$D^0D^{\ast0}$ threshold, for values of $\lambda$ close to 3. The two
bullets and the star correspond to the pole positions $1,2,3$ for which the
$D^0D^{\ast0}$ amplitude is depicted in the right-hand plot. All three cases
are compatible with the data \cite{PRL91p262001}, though pole no.~3 agrees best
with the PDG value of $(3871.57\pm0.25)$~MeV \cite{PDG2010} for the $X(3872)$
mass. The inset in the latter plot compares the amplitudes in the OZI-forbidden
channels $\omega\jpsi$ and $\rho^0\jpsi$, which are coupled to the $X(3872)$
system, too. Although their influence on the pole positions is almost
negligible, owing to their small couplings, they must nevertheless be included
for a realistic description of the $\pi^+\pi^-\pi^0\jpsi$ and $\pi^+\pi^-\jpsi$
data, respectively. What the inset shows is that the $\omega\jpsi$ and
$\rho^0\jpsi$ amplitudes are of comparable size, at least at the energy of the
pole (no.~2 in this case), despite the fact that the $\omega\jpsi$ coupling is
much larger than that of the isospin-breaking $\rho^0\jpsi$ channel. Note that
the latter channels are effectively smeared out in order to account for the
widths of the $\omega$ and especially the very broad $\rho$. This is done using
a novel unitarisation procedure for an $S$-matrix with complex masses in the
asymptotic states \cite{EPJC71p1762}. Thus, the $\rho^0\jpsi$ decay mode
gets enhanced, such that
$\Gamma(\omega\jpsi)/\Gamma(\rho^0\jpsi)\sim1$, in agreement with
data \cite{PDG2010}.

\section{Understanding the axial-vector charmed mesons}
One of the major puzzles in open-charm spectroscopy is the seemingly odd
pattern of masses and widths of the axial-vector (AV) charmed mesons \dc, \dd,
\dsc, and \dsd\ \cite{PDG2010}. In the first place, the two $c\bar{q}$
($q=u,d$) states are almost degenerate in mass, while there is a 76 MeV mass
splitting between the two $c\bar{s}$ AV mesons. Secondly, the \dc\ and \dsc\
resonances are much narrower than what one would naively expect from their
OZI-allowed $S$-wave decay modes $D^\ast\pi$ and $D^\ast K$, respectively.
Finally, both $c\bar{s}$ states are very narrow (0--3~MeV), whereas the \dd\
is very broad ($\sim\!400$~MeV) and the \dc\ has a modest width (20--25~MeV).
Standard quark models as well as chiral-Lagrangian approaches to heavy-light
mesons completely fail to predict such a pattern \cite{PRD84p094020}.

\begin{figure}[!hb]
\begin{tabular}{cc}
\hspace*{1cm}
\epsfig{file=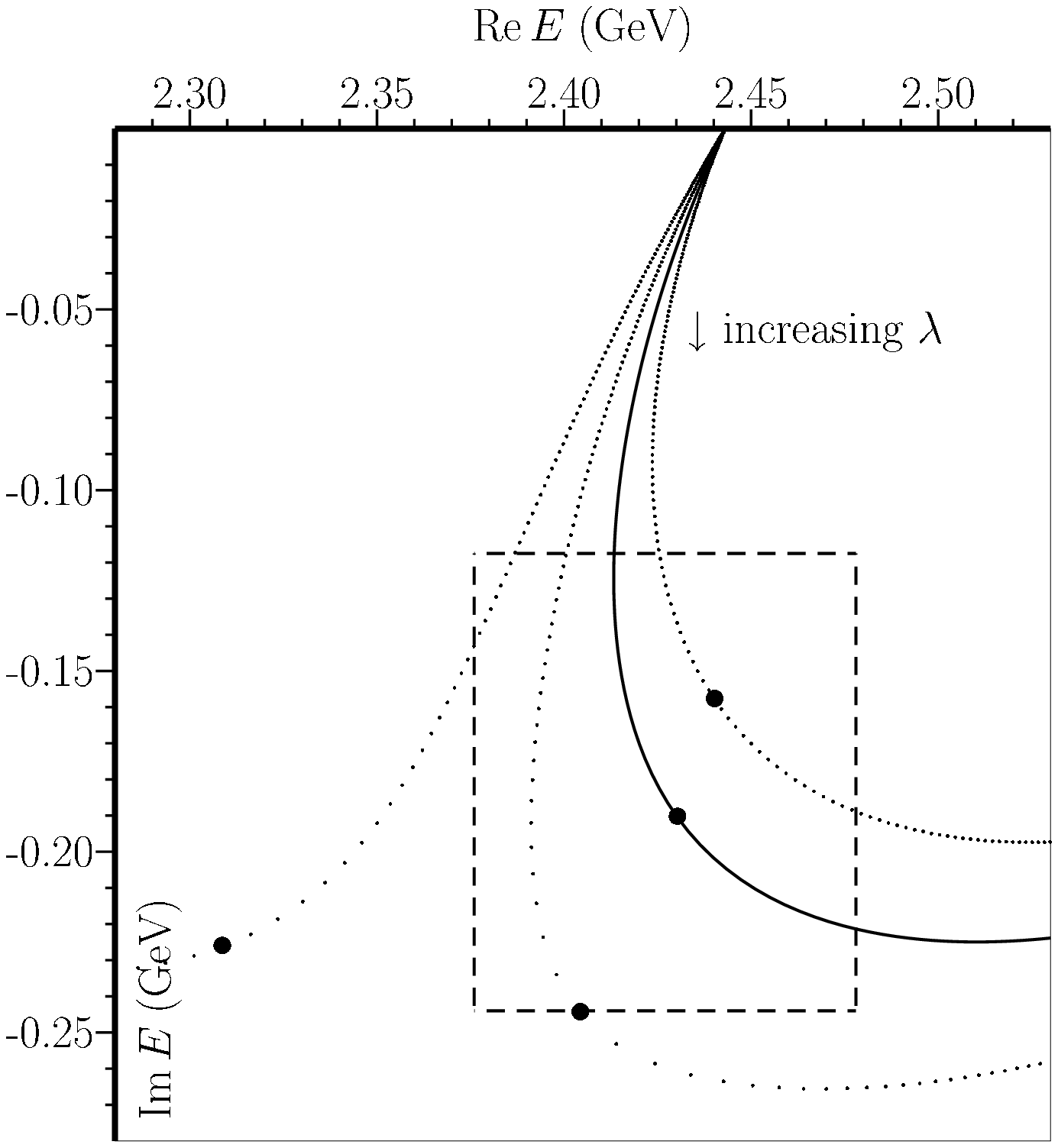,scale=0.5} &
\hspace*{1cm}
\epsfig{file=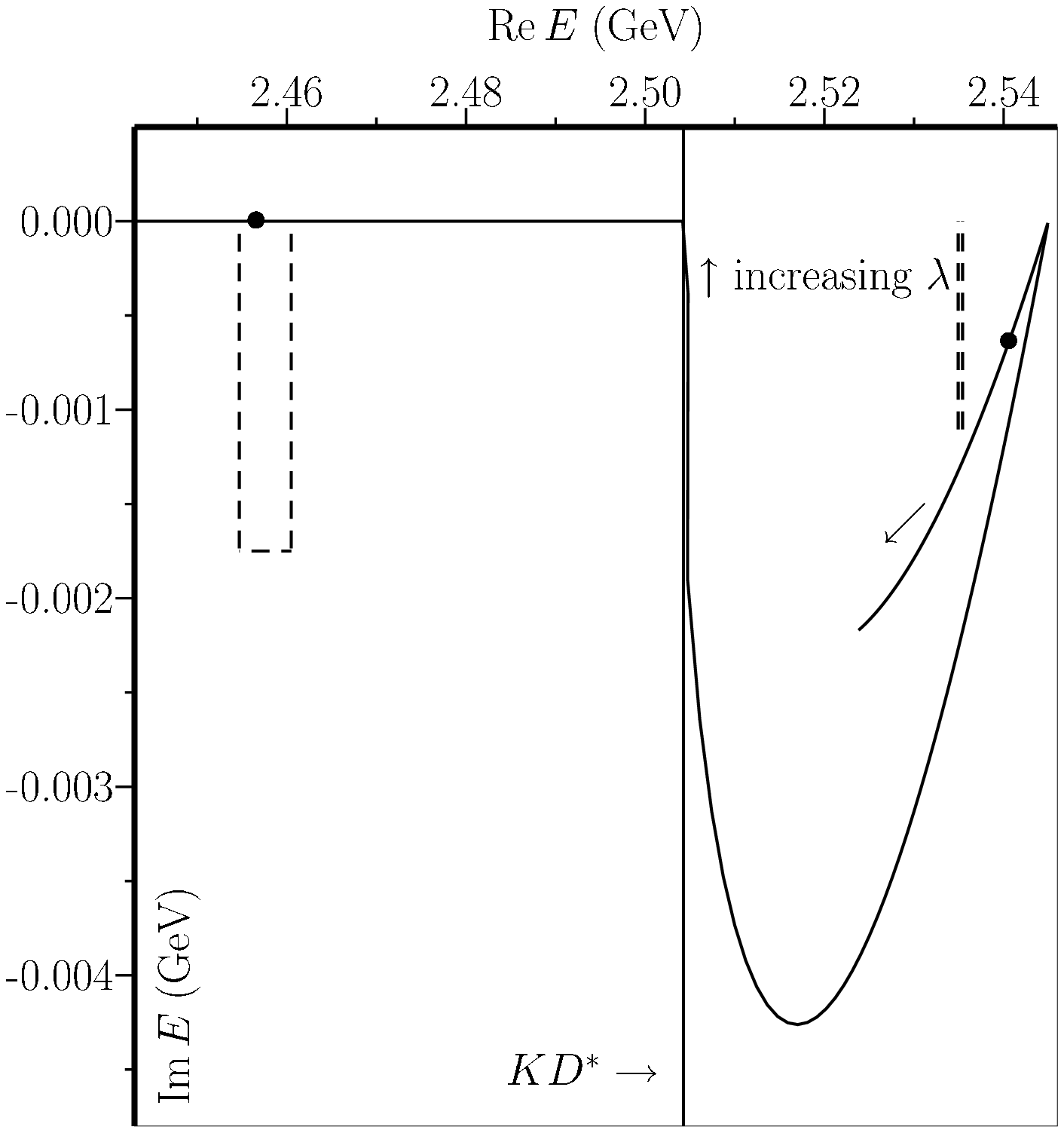,scale=0.5} 
\end{tabular}
\caption{Left: pole trajectories of \dd\ as a function of $\lambda$, for
$r_0=$ 3.2--3.5~GeV$^{-1}$;
right: trajectories of \dsc\ and \dsd, as a function of $\lambda$, for
$r_0=3.12$~GeV$^{-1}$.
Left and right: dashed box, rectangle, and strip delimit experimental
\cite{PDG2010} \dd, \dsd, resp.\ \dsc\ pole. Also see text.}
\label{axialcharm}
\end{figure}

In Ref.~\cite{PRD84p094020}, we tackled the AV charmed mesons by coupling
bare $c\bar{q}$ and $c\bar{s}$ states, each with a \tpo\ and a \spo\ component,
to the relevant OZI-allowed meson-meson channels, most importantly $D^\ast\pi$
for $c\bar{q}$
and $D^\ast K$ for $c\bar{s}$. Notice that charmed mesons have no definite 
$C$-parity, so that both \tpo\ ($J^{PC}=1^{++}$) and \spo\ ($J^{PC}=1^{+-}$)
contribute to the wave function. In Fig.~\ref{axialcharm}, we show how the
bare $c\bar{q}$ and $c\bar{s}$ states at 2443 and 2545~MeV \cite{PRD84p094020},
respectively, get shifted and split up because of the coupled meson-meson
channels. However, in either case, one pole moves very little, thus behaving
as a quasi-bound state in the continuum \cite{PRD84p094020}, whereas the other
shifts a lot. Note that the relatively narrow \dc, with pole at
$(2439-i3.5)$~MeV, is not shown in Fig.~\ref{axialcharm}, for clarity. Also,
the bullets correspond to the optimum coupling values $\lambda=1.30$ (1.19)
for $c\bar{q}$ ($c\bar{s}$). As a result, the \dd\ becomes very broad, but
with practically the same central mass as the \dc.
On the other hand, the \dsd\ pole does move to lower energies, even below the
$D^\ast K$ threshold, in agreement with experiment. Overall, four masses and
four widths are quite well reproduced with only two parameters, which moreover
have little freedom. Moreover, the mixing angle of about $35^\circ$ between the
\tpo\ and \spo\ components, needed to describe the data, is generated here
fully dynamically and not put in by hand. For further details, see
Ref.~\cite{PRD84p094020}. 

\section{Surprise in BABAR data: a very light scalar of 38 MeV?}
\begin{figure}[!h]
\begin{tabular}{c}
\resizebox{!}{140pt}{\includegraphics{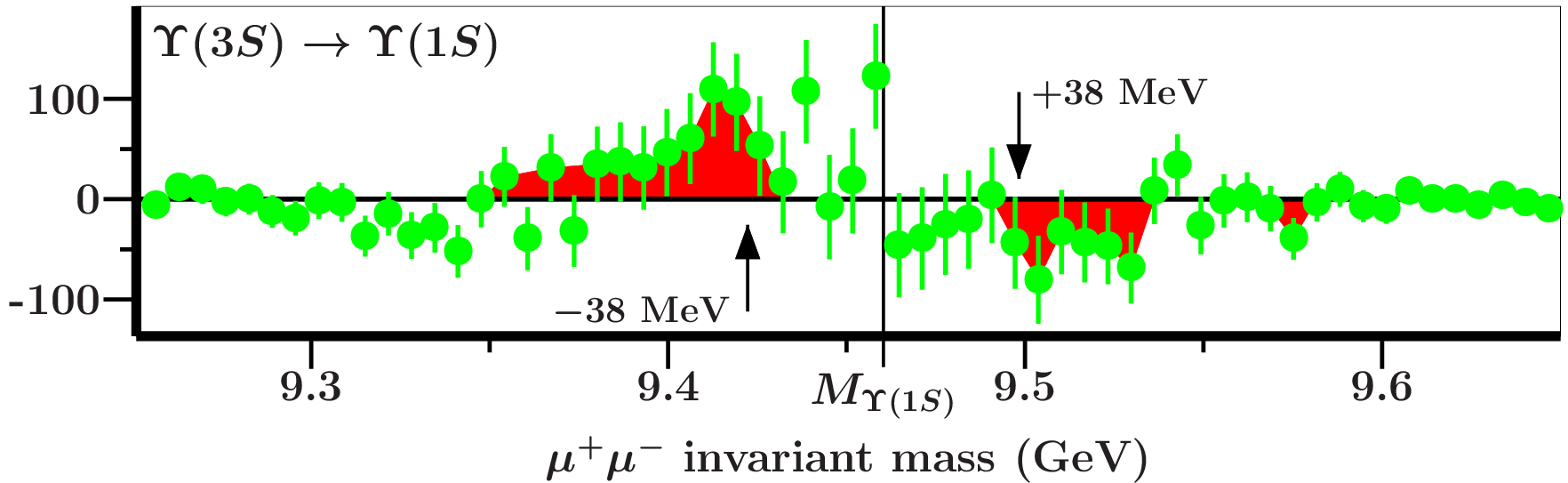}}
\\[6mm]
\hspace*{16.5mm}
\resizebox{!}{113pt}{\includegraphics{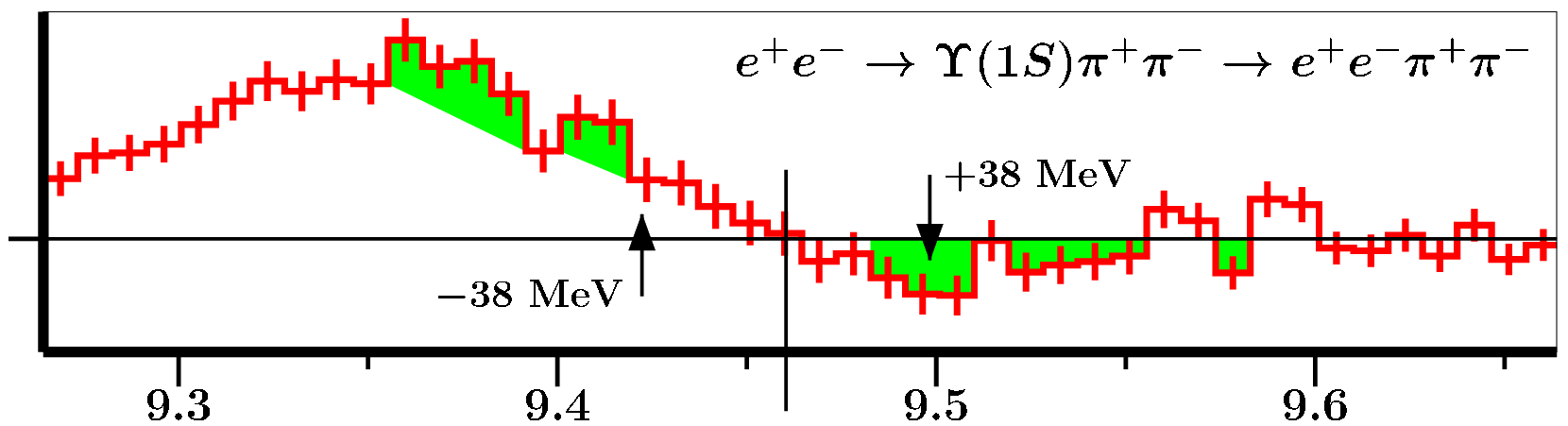}}
\end{tabular}
\caption{Top: event distribution of the excess signal \cite{HEPEX09100423}
in the invariant $\mu^{+}\mu^{-}$ mass 
for the reaction $\Upsilon\left( 3\,{}^{3\!}S_{1}\right)$ $\to$
$\pi^{+}\pi^{-}\Upsilon\left( 1\,{}^{3\!}S_{1}\right)$
$\to$ $\pi^{+}\pi^{-}\mu^{+}\mu^{-}$,
using bins of 6.5 MeV. 
Bottom: excess signals in invariant $e^{+}e^{-}$ mass distributions
for the reaction
$e^{+}e^{-}\to\pi^{+}\pi^{-}\Upsilon\left( 1\,{}^{3\!}S_{1}\right)$
$\to\pi^{+}\pi^{-}e^{+}e^{-}$,
using all available data \cite{PRD78p112002} and bins of 9 MeV;
statistical errors are shown by vertical bars.}
\label{e38}
\end{figure}

Observation of the $\Upsilon$(\ttdo) \cite{HEPPH10094097} and seeing
indications of the $\Upsilon$(\otdo) \cite{HEPPH10094097} in published
\cite{PRD78p112002} BABAR data stimulated further analysis of these
data. Surprisingly, evidence was found \cite{HEPPH11021863} for the existence
of a very light scalar boson of strong interactions, with a mass of about
38~MeV. This was based on clear and structured excess signals in
invariant-mass projections of $e^+e^-$ and $\mu^+\mu^-$ pairs for the reactions
$e^{+}e^{-}$ $\to$ $\pi^{+}\pi^{-}\Upsilon\left( 1,2\,{}^{3\!}S_{1}\right)$
$\to$ $\pi^{+}\pi^{-}e^{+}e^{-}$ (and $\to\pi^{+}\pi^{-}\mu^{+}\mu^{-}$).
In fact, similar excess signals were observed, though not explained, in
Ref.~\cite{HEPEX09100423}, for the process
$\Upsilon(3S)\to\pi^{+}\pi^{-}\Upsilon(1S)\to\pi^{+}\pi^{-}\mu^{+}\mu^{-}$,
as depicted here in Fig.~\ref{e38}, upper plot.
In the lower plot, we show very similar signals, with the same structure, in 
$e^{+}e^{-}$ $\to$ $\pi^{+}\pi^{-}\Upsilon(1S)\to\pi^{+}\pi^{-}e^{+}e^{-}$ 
data \cite{PRD78p112002}, which clearly suggest emission of a light quantum.
For discussion, theoretical justifications, and more experimental evidence
supporting a very light scalar for strong interactions, see
Ref.~\cite{HEPPH11021863}.

\section{Conclusions}
In the foregoing we have shown, at the hand of several examples of puzzling
mesonic structures, that genuine meson spectroscopy is much more involved 
than just hunting after the biggest bump in data and trying to reproduce
its central mass with some confining potential. In the case of true resonances,
like the $X(3872)$ and the AV charmed mesons \dc, \dd, \dsc, and \dsd, large
and non-linear effects are found from unitarisation, which makes a reliable
determination of their masses and widths a complicated task. For non-resonant
enhancements like the $X(4260)$ --- in our view --- a very careful analysis
of the available data is required, as well as accounting for competing
channels and existing resonances in the same energy region. We believe that
the RSE, both for scattering, as described in Sec.~3, and for production
processes \cite{AOP323p1215}, is a very powerful yet simple-to-use formalism,
which can be adapted to any quark potential model. 

Unquenching is becoming more and more mandatory in quark models, just like in
lattice QCD. To think that its effects can be neglected in spectroscopy, or
mimicked by a global redefinition of quark masses, is an illusion.
\section*{Acknowledgements}
Two of us (EvB, GR) thank J.~Segovia for collaboration on the $X(4260)$
(Sec.~2).
This work was supported in part by the {\it Funda\c{c}\~{a}o para a
Ci\^{e}ncia e a Tecnologia} \/of the {\it Minist\'{e}rio da Ci\^{e}ncia,
Tecnologia e Ensino Superior} \/of Portugal, under contract
CERN/FP/116333/2010 and grant SFA-2-91/CFIF.


\begin{thebibliography}{99}
\itemsep -2pt 
\bibitem{PDG2010}
K.~Nakamura {\it et al.}  [Particle Data Group],
\Journal{\JPG}{37}{075021}{2010}and
2011 online update

\bibitem{PoSHQL2010p003}
E.~van Beveren and G.~Rupp,
{\it PoS HQL2010} \/(2010) 003 
[arXiv:1011.2360 [hep-ph]]

\bibitem{HEPEX11054583}
I.~Adachi {\em et al.} \/[Belle Collaboration],
arXiv:1105.4583 [hep-ex]

\bibitem{EPL96p11002}
D.~V.~Bugg,
\Journal{\EPL}{96}{11002}{2011}
[arXiv:1105.5492 [hep-ph]]

\bibitem{HEPPH11061552}
I.~V.~Danilkin, V.~D.~Orlovsky, and Yu.~A.~Simonov,
arXiv:1106.1552 [hep-ph]

\bibitem{ZPC30p615}
E.~van Beveren, T.~A.~Rijken, K.~Metzger, C.~Dullemond, G.~Rupp,
and J.~E.~Ribeiro,
\Journal{\ZPC}{30}{615}{1986}
[arXiv:0710.4067 [hep-ph]]

\bibitem{PLB641p265}
E.~van Beveren, D.~V.~Bugg, F.~Kleefeld, and G.~Rupp,
\Journal{\PLB}{641}{265}{2006}
[arXiv:hep-ph/0606022]

\bibitem{PRL95p142001}
B.~Aubert {\it et al.}  [BABAR Collaboration],
\Journal{\PRL}{95}{142001}{2005}
[arXiv:hep-ex/0506081]

\bibitem{PRL105p102001}
E.~van Beveren, G.~Rupp, and J.~Segovia,
\Journal{\PRL}{105}{102001}{2011}
[arXiv:1005.1010 [hep-ph]]

\bibitem{AOP324p1620}
E.~van Beveren and G.~Rupp,
\Journal{\AOP}{324}{1620}{2009}
[arXiv:0809.1149 [hep-ph]]

\bibitem{AOP323p1215}
E.~van Beveren and G.~Rupp,
\Journal{\AOP}{323}{1215}{2008}
[arXiv:0706.4119 [hep-ph]]

\bibitem{PRL91p262001}
S.~K.~Choi {\it et al.}  [Belle Collaboration],
\Journal{\PRL}{91}{262001}{2003}
[arXiv:hep-ex/0309032]

\bibitem{EPJC71p1762}
S.~Coito, G.~Rupp, and E.~van Beveren,
\Journal{\EPJC}{71}{1762}{2011}
[arXiv:1008.5100 [hep-ph]]

\bibitem{PRD84p094020}
S.~Coito, G.~Rupp and E.~van Beveren,
\Journal{\PRD}{84}{094020}{2011}
[arXiv:1106.2760 [hep-ph]]

\bibitem{HEPPH10094097}
E.~van Beveren and G.~Rupp,
arXiv:1009.4097 [hep-ph]

\bibitem{PRD78p112002}
B.~Aubert {\it et al.} [BABAR Collaboration],
\Journal{\PRD}{78}{112002}{2008}
[arXiv:0807.2014 [hep-ex]]

\bibitem{HEPPH11021863}
E.~van Beveren and G.~Rupp,
arXiv:1102.1863 [hep-ph]

\bibitem{HEPEX09100423}
E.~Guido [BABAR Collaboration],
arXiv:0910.0423 [hep-ex]

\end{thebibliography}
\end{document}